\documentstyle[preprint,aps]{revtex}
\begin{document}

\hsize\textwidth\columnwidth\hsize\csname@twocolumnfalse\endcsname
\title{N=2-Supersymmetric Dynamics of a Spin-1/2 Particle in an Extended External Field}
\author{G. S. Dia$s^1${\thanks {E-mail - dias@cbpf.br }} and 
J. A. Helay\"{e}l-Net$o^{1,2}${\thanks {E-mail - helayel@gft.ucp.br }}} 
\address{Centro Brasileiro de Pesquisas F\'\i sica$s^1$ \\ Rua Xavier
Sigaud, 150 Rio de Janeiro, RJ 22290-180, Brasil
\\Grupo de F\'{\i}sica Te\'orica, GFT-UC$P^2$ \\ Rua Bar\~ao do Amazonas, 124
Petr\'{o}polis, RJ 25685-070 , Brasil} 
\maketitle
\begin{abstract} 
One considers the quantum dynamics of a charged spin-\( \frac{1}{2} \) particle
in an extended external eletromagnetic field that arises from the reduction
of a 5-dimensional Abelian gauge theory. The non-relativistic regime
of the reduced 4D-dynamics is worked out and one identifies the system as a sector
of an N=2-supersymmetric quantum-mechanical dynamics. The full supersymmetric
model is studied and one checks the algebra of the fermionic charges; 
the conclusion is that no central charge drops out. The possible r\^ole
of the extra external fields, a scalar and a magnetic-like field, is discussed.
\end{abstract} 
\pacs{PACS:}  
\vskip1pc \narrowtext

\section{Introduction}

Supersymmetric quantum field theory at low energies (non-relativistic limit)  
raises special interest for a great deal of reasons, \cite{monopolosusyspinB}, 
\cite{aplicacaosusy}, \cite{capitulolivrosusy}, \cite{shapeinvariance}. 
If Supersymmetry is a 
symmetry of Nature, what we see today must be its low-energy remnant through 
the breaking realised by means of some mechanism \cite{susyvinculosquebra}, 
\cite{spinsusyB}, \cite{susyquebramecanicaquantica}. In this limit, the underlying 
field theory should approach a Galilean invariant supersymmetric field theory and, 
by the Bargmann superselection rule \cite{TQCanposvistacomMquantica}, such a field
theory should be equivalent to a supersymmetric Schr\"{o}dinger equation in
each particle number sector of the theory.

In this paper, we shall consider a particle of mass m and spin-1/2 in an
external field, \cite{monopolospinsusyA}, \cite{monopolosusyC}, 
\cite{ideiainicial},\cite{spinemsusy}. We build up a superspace
action for this model, read off the supersymmetry. transformations of the 
component coordinates and then obtain the supersymmetric charge operators 
\cite{calculocarga}. We analyse their algebra and pursue
the investigation of the possible existence of a central charge in the system.

We set our discussion by first considering the case of (1+2) dimensions. 
The Dirac action in such a space-time may be written as

\[
L=\overline{\Psi }(i\gamma ^{\mu }D_{\mu }-m)\Psi ,
\]
where $\eta =(1,-1,-1)$ and $\gamma ^{0}=\sigma _{3}$, $\gamma ^{1}=i\sigma
_{2}$, $\gamma ^{3}=\sigma _{1}$, $\gamma _{3}=i\gamma _{0}\gamma _{1}$, $
D_{\mu }\equiv\partial _{\mu }+ieA_{\mu }$. 
From the equation of motion,

\[
(i\gamma ^{\mu }D_{\mu }-m)\Psi =0,
\]

\[
\Psi =\left( 
\begin{array}{c}
\psi _{1} \\ 
\psi _{2}
\end{array}
\right), 
\]
and with the approximation 

\[
E+m-e\phi\cong 2m, 
\]

\[
\psi _{2}\cong -\frac{(D_{3}+iD_{1})}{2m}\psi _{1}, 
\]
the Pauli Hamiltonian in 1+2 \cite{susydimensaodois} takes over the form:

\[
H=\frac{1}{2m}\overrightarrow{\nabla }^{2}-\frac{ie}{2m}\overrightarrow{
\nabla }.\overrightarrow{A}-\frac{ie}{m}\overrightarrow{A}.\overrightarrow{
\nabla }-\frac{e^{2}}{2m}\left| \overrightarrow{A}\right| ^{2}-e\phi. 
\]
The minimal coupling does not generate here any coupling term of the type
spin-magnetic field as it is the case for the Pauli Hamiltonian in D=1+3. 
We can also show that this coupling term cannot 
appear from a non-minimal coupling of the type:

\[
D_{\mu }=\partial _{\mu }+ieA_{\mu }+ig\widetilde{F}_{\mu } ,
\]

\[
\widetilde{F}_{i}=\frac{1}{2}\epsilon _{ijk}F^{jk} .
\]

In (1+3)D, we see that the bosonic degree of freedom, $x^{i}(t)$ (position), matches with 
the fermionic degree of freedom, $S^{i}(t)$ (spin). This makes possible the 
construction of superfields and then of a manifestally supersymmetric Lagrangian 
with linearly realised supersymmetry \cite{calculocarga}. However, in (1+2)D, the same 
does not occur; angular momentum is not a vector any longer, the number of degrees
of freedom of $x^i$ and the spin do not match anymore and then we cannot in this way 
build up superfields nor a supersymmetric Lagrangian in superspace. 
We can try to work in (1+3)D, starting from Dirac, taking the non-relativistic limit 
and making a dimensional reduction to (1+2)D. But, we see that it is not
possible to drag into (1+2)D the analogue of the coupling of spin-magnetic field
that exists in (1+3)D \cite{susydimensaodois}. We will try to study this model in 
(1+3)D with N=1 and N=2 supersymmetry; we shall however start from (1+4)D, where
we may propose a matching between coordinates and 5D spin and then carry out the 
dimmensional reduction to 4D, getting to an extended eletromagnetic-like field.

\section{N=1-D=5 Reduced to N=1-D=4}

We consider the following action in D=4+1,

\begin{equation}  \label{Ldirac5d}
L=\overline{\Psi }(i\Gamma \hat{^{\mu }}D_{\hat{\mu }}-m)\Psi ;
\end{equation}

where we define: 
\begin{equation}  \label{Definicao1}
\hat{\mu }\in \{0,1,2,3,4\}, \quad \eta =(+,-,-,-,-), \quad
\end{equation}

\[
x\hat{^{\mu }}=(t,x,y,z,w), \quad F_{\hat{\mu }\hat{\nu }}=\partial _{\hat{%
\mu }}A_{\hat{\nu }}- \partial _{\hat{\nu }}A_{\hat{\mu }}, 
\]

\[
D_{_{\hat{\mu }}}=\partial _{\hat{\mu }}+ieA_{\hat{\mu }},\quad A_{\mu
}=(A_{i},\Lambda ), \quad \left\{ \Gamma ^{\hat{\mu }}, \Gamma ^{\hat{\nu }%
}\right\} =2\eta ^{\hat{\mu }\hat{\nu }}, 
\]

\[
\partial ^{i}\Longleftrightarrow \overrightarrow{\nabla },\quad \partial
_{0}\Longleftrightarrow \frac{\partial }{\partial t},\quad
E^{i}\Longleftrightarrow \overrightarrow{E},\quad B^{i}\Longleftrightarrow 
\overrightarrow{B},\quad 
\]

\[
F_{04}={\cal {E}}, \quad F_{4i}={\cal {B}}_{i},\quad F_{ij}=\epsilon
_{ijk}B_{k}, \quad F_{0i}=E_{i}, 
\]
where $i,j,k\in \{1,2,3\} $. Notice that the scalar, ${\cal {E}}$, and the vector,
$\overrightarrow{\cal{B}}$, accompany the vectors $\overrightarrow{E}$ and 
$\overrightarrow{B}$, later on to be identified with the electric and magnetic fields,
respectively. Our explicit representation
for the gamma-matrices is given in the Appendix. We set:
\[\Psi =\left( 
\begin{array}{c}
\psi _{1} \\ 
\psi _{2} \\ 
\psi _{3} \\ 
\psi _{4}
\end{array}
\right) , \quad \overline{\Psi }=\Psi ^{\dagger }\Gamma ^{0} \]

We now take the equation of motion for ${\Psi }$ from the
Euler-Lagrange equations (\ref{Ldirac5d}). We suppose a stationary
solution, 
\[
\Psi (x,t)=\exp (i\epsilon t)\chi (x); 
\]
here, we are considering that the external field does not depend on t, and 
$A_{0}=0$. Then, proposing the reduction prescription:

\[
\partial _{4}(A_{\mu })=0, 
\]
we obtain

\begin{equation}  \label{definicao2}
\overrightarrow{E}\Longleftrightarrow F_{0i}\equiv \partial _{0}A_{i}
-\partial _{i}A_{0}=0;
\end{equation}

\[
{\cal {E}}\equiv F_{04}\equiv \partial_{0} A_{4}-\partial_{4}A_{0}=0; 
\]

\[
F_{ij}=\partial _{i}A_{j}-\partial _{j}A_{i}\equiv -\epsilon _{ijk}B_{k}; 
\]

\[
F_{i4}=\partial _{i}A_{4}-\partial _{4}A_{i}\equiv \partial _{i}\Lambda
\equiv -{\cal {B}}_{i}.
\]

Then, considering the non-relativistic limit of the reduced theory, 
the equations of motions read:

\begin{equation}
(\epsilon +m)I_{_{2}}\chi _{1}+(-i\sigma ^{i}(\partial
_{i}+ieA_{i})+iI_{_{2}}eA_{4})\chi _{2}=0  \label{eqespinorial}.
\end{equation}
\[
(i\sigma ^{i}(\partial _{i}+ieA_{i})+iI_{_{2}}eA_{4})\chi _{1}+(-\epsilon
+m)I_{_{2}})\chi _{2}=0. 
\]
therefore, we see that $\chi=\left(\begin{array}{c}
\chi _{1} \\ 
\chi _{2}
\end{array}
\right) $ looses two degrees of fredom, described by the "weak" spinor, $\chi _{2}$.
So the Pauli-like Hamiltonian we get to reads as below:

\begin{equation}  \label{Hpauli54dn1}
H=-\frac{1}{2m} \left[ \left( \overrightarrow{\nabla } -ie\overrightarrow{A}%
\right) ^{2} +e\overrightarrow{\sigma }\left( \overrightarrow{\nabla }\times 
\overrightarrow{A}\right) \right.
\end{equation}

\[
\left. -e\overrightarrow{\sigma }\overrightarrow{\nabla }(A_{4})
-e^{2}(A_{4})^{2}\right],
\]
where we used (\ref{Definicao1}). 

If we define, $P^{i}=-i\overrightarrow{\nabla } ^{i}$ ($\hbar=1$), then the
Hamiltonian (\ref{Hpauli54dn1}) will become

\[H=\frac{1}{2m}\left[ \left( p_{i}-ieA_{i}\right)
^{2}-eB_{i}S_{i} +e{\cal {B}}_{i}S_{i}-\frac{e^{2}}{2}\Lambda ^{2}\right] ; \]
the corresponding Lagrangian is given by:

\begin{equation}  \label{Lpauli54dn1}
L=\frac{1}{2}(\dot{x}_{i})^{2}+\frac{i}{2}\psi _{i}\dot{\psi }_{i} +eA_{i}
\dot{x}_{i}+
\end{equation}
\[
\frac{ie}{2}B_{i}\epsilon _{ijk}\psi _{j}\psi _{k} -\frac{ie}{2}{\cal {B}}
_{i}\epsilon _{ijk}\psi _{j}\psi _{k} -\frac{e^{2}\Lambda ^{2}}{2}, 
\]
where the dot is a derivative with respect to  t. We can also write

\begin{equation}  \label{Lpauli54dn1spin}
L=\frac{1}{2}(\dot{x}_{i})^{2}+\frac{i}{2}\psi _{i}\dot{\psi }_{i}+eA_{i}
\dot{x}_{i},
\end{equation}
\[
-eB_{i}S_{i}+e{\cal {B}}_{i}S_{i}-\frac{e^{2}\Lambda ^{2}}{2}, 
\]
where we define the spin by the product below:

\[
S_{i}=-\frac{i}{2}\epsilon _{ijk}\psi _{j}\psi _{k}. 
\]

\section{N=1-Supersymmetric action}

 To render more systematic our discussion, we think it is advisable to 
 set up a superfield approach. We can define the N=1-supersymmetric model
 in analogy with the model presented above, eq.(\ref{Lpauli54dn1spin}). 
 We start defining the superfelds by

\begin{equation}
\Phi _{i}(t,\theta )=x_{i}(t)+i\theta \psi _{i}(t)\quad \Sigma (t,\theta
)=\xi (t)+\theta R(t),  \label{Definicao3}
\end{equation}

\[
\Lambda (x)=A_{4}(x) \quad \Lambda (\Phi _{s})=\Lambda (x) +i(\partial
_{j}\Lambda (x))\theta \psi _{j}. 
\]

The supercharge operators and the covariant derivatives are given by: 
\begin{equation}
Q=\partial _{\theta }+i\theta \partial _{t}\quad D=\partial _{\theta
}-i\theta \partial _{t}\quad H=i\partial _{t}.  \label{Operadoresn1}
\end{equation}

Then, the N=1-supersymmetric Lagrangian that generates, Lagrangian 
(\ref{Lpauli54dn1}) can be written in superfields as: 
\begin{equation}
{\cal {L}}=\frac{i}{2}\dot{\Phi}_{i}D\Phi _{i}+ie(D\Phi _{i})A_{i}(\Phi )+%
\frac{1}{2}\Sigma D\Sigma +  \label{Lsusyn1}
\end{equation}
\[
-e\Sigma \Lambda (\Phi )+\frac{ie}{2}\epsilon _{ijk}\partial _{i}\Lambda
(\Phi )\Phi _{j}D\Phi _{k}, 
\]
where we have set $m=1$.

The  commutators and anticommutators for the superfield
components are

\begin{equation}  \label{Relacaocomutacaon1}
\left[ \psi _{i,}\psi _{j}\right] _{+}= \delta _{ij},\quad \left[ \xi , \xi
\right] _{+}=-2i,\quad \left[ x_{i,}p_{j}\right] =i\delta _{ij},
\end{equation}

\[
\left[ \psi _{i,}\xi \right] _{+}=0; 
\]
the other commutators vanish.

The component field R does not have dynamics; then, we use the equation of
motion to remove it from the Lagrangian:
\begin{equation}
R=e\Lambda  \label{Eqr}.
\end{equation}

The supersymmetric action is then 
\[
{\cal {S}}=\int dtd\theta {\cal {L}} 
\]

We have written the supersymmetric Lagrangian in components, \cite{calculocarga}, 
under the form,

\[
{\cal {L}}=K+\theta L 
\]
where we define $K$ and $L$ as follows:

\[
K=-\frac{1}{2}\dot{x}_{i}\psi _{i}-e\psi _{i}A^{i}+\frac{\xi R}{2}-e\xi
\Lambda -\frac{e}{2}\epsilon _{ijk}{\cal {B}}_{i}x_{j}\psi _{k} 
\]

\begin{equation}  \label{Lsusycomponentesn1}
L=\frac{1}{2}\dot{x}_{i}\dot{x}_{i}- \frac{1}{2}i\dot{\psi }_{i}\psi _{i} +e%
\dot{x}_{i}A_{i}-\frac{ie}{2}F_{ij}\psi _{i}\psi _{j}+
\end{equation}

\[
+\frac{i}{2}e\xi \dot{\xi }+\frac{R^{2}}{2}-eR\Lambda +ie\xi \psi _{j}{\cal {%
B}}_{j}+ 
\]
\[
+\frac{e}{2}\epsilon _{ijk}{\cal {B}}_{i}x_{j}\dot{x}_{k} -\frac{ie}{2}%
\epsilon _{ijk}{\cal {B}}_{i}\psi _{j}\psi _{k} -\frac{ie}{2}\epsilon
_{ijk}\psi _{r}\partial _{i} {\cal {B}}_{r}x_{j}\psi _{k} 
\]

Also from $L$ in (\ref{Lsusycomponentesn1}), we can read the Hamiltonian: 

\begin{equation}  \label{Hsusycomponentesn1}
H=\frac{1}{2}(p^{i}-eA^{i})^{2} +\frac{ie}{2}\epsilon _{ijk}B_{k}\psi
_{i}\psi _{j}+
\end{equation}

\[
-\frac{ie}{2}\epsilon _{ijk}{\cal {B}}_{i}\psi _{j}\psi _{k} -\frac{ie}{2}%
\epsilon _{rji}(\partial _{r}{\cal {B}}_{k})x_{j} \psi _{k}\psi _{i}+ie 
{\cal {B}}_{i}\xi\psi _{i}+ 
\]

\[
-\frac{e}{2}\epsilon _{jki}{\cal {B}}_{j}x_{k}(p_{i}-eA^{i}) +\frac{e^{2}}{8}%
\epsilon _{jki}\epsilon _{rni}{\cal {B}}_{j}{\cal {B}}_{r}x_{k}x_{n}+\frac{%
e^{2}\Lambda ^{2}}{2}; 
\]
in this equation, we have eliminated the component-field without dynamical character.

\subsection{The supersymmetric charge}

Acting with the supercharge operator (\ref{Operadoresn1}) on the superfields 
(\ref{Definicao3}), we can obtain the supersymmetry transformations of 
components fields; they are:

\begin{equation}  \label{Transfn1}
\delta \psi ^{i}=\epsilon \dot{x}^{i},\quad \delta \xi =-i\epsilon R,\quad
\delta R=-\epsilon \dot{\xi },\quad \delta x^{i}=-i\epsilon \psi ^{i}.
\end{equation}

From the supersymmetric transformations and the Lagrangian (\ref
{Lsusycomponentesn1}), we can analytically calculate the supercharge,
through the Noether's theorem. The charge operator comes out to be

\begin{equation}  \label{Cargaanalitican1}
Q=\psi _{i}\dot{x}_{i}+\frac{1}{2}(1-i)\xi R- e\xi \Lambda.
\end{equation}

The supercharge algebra reads

\[
\left[ Q,Q\right] _{+}=2H \quad $or$ \quad Q^{2}=H, 
\]
where H is the Hamiltonian of eq. (\ref{Hsusycomponentesn1}).

\subsection{Equations of motion for the extended Electromagnetism in D=4+1}

We start by considering the equations of motion

$\epsilon ^{\hat{\mu}\hat{\rho}\hat{\sigma}\hat{\beta}\hat{\gamma}}\partial
_{\hat{\sigma}}F_{\hat{\beta}\hat{\gamma}}=0$ and $\partial _{\hat{\sigma}
}F^{\hat{\sigma}\hat{\gamma}}=\rho ^{\hat{\gamma}}$; where 
$F_{\hat{\beta}\hat{\gamma}}$ is the field strength defined in 
(\ref{Definicao1}) and 
$\epsilon^{\hat{\mu}\hat{\rho}\hat{\sigma}\hat{\beta}\hat{\gamma}}$ is 
the Levi-Civita tensor in 5 dimensions. Then, we have:

\begin{eqnarray}
\label{Eqeletro5d}
\overrightarrow{\nabla }\cdot \overrightarrow{{\cal {B}}}=J^{5}, \nonumber \\ 
\overrightarrow{\nabla }\times\overrightarrow{{\cal{B}}}=0, \nonumber \\
\overrightarrow{\nabla }\cdot\overrightarrow{B}=0, \nonumber \\
\overrightarrow{\nabla }\times \overrightarrow{E}=-\frac{\partial
\overrightarrow{B}}{\partial t}, \\
\overrightarrow{\nabla }\cdot \overrightarrow{E}=J^{0}, \nonumber \\
\overrightarrow{\nabla }\cdot {\cal {E}}=-\frac{\partial \overrightarrow{
{\cal {B}}}}{\partial t} \nonumber,\\
\overrightarrow{\nabla } \times 
\overrightarrow{B}=
\frac{\partial\overrightarrow{E}}{\partial t}+J^{i}. \nonumber 
\end{eqnarray}

The Lorentz force, calculated using the fact that the
Hamiltonian in D=4+1 is

\begin{equation}  \label{Hpauli5d}
H=\frac{1}{2m}\left( \overrightarrow{p} +e\overrightarrow{A}\right)
^{2}+eA^{0},
\end{equation}
reads as below:

\begin{equation}  \label{ForcaLorentz5d}
m\frac{\partial ^{2}x^{5}}{\partial t^{2}}=e\overrightarrow{{\cal {B}}} 
\overrightarrow{v}+e{\cal {E}},
\end{equation}

\[
m\frac{\partial ^{2}\overrightarrow{x}}{\partial t^{2}}=e\overrightarrow{v}
\times \overrightarrow{B}+e\overrightarrow{{\cal {B}}}\dot{x}_{5} +e\overrightarrow{E}. 
\]

\section{A toy Model with N=2-su.sy.}

The superspace coordinates are $(t,\theta _{1},\theta _{2})$ with 
$\theta _{1}$ and $\theta _{2}$ real, or $(t,\theta,\bar{\theta})$ with 
$\theta =\theta_{1}+i\theta _{2}$ and $\bar{\theta}$ is it complex conjugate. 
$\theta $, $\bar{\theta}$ are complex Grasmannian variables.

The translations 

\[\delta t=i\epsilon^{*}\theta^{*}+i\epsilon\theta, \quad 
\delta\theta^{*}=i\epsilon^{*}\quad and \quad \delta\theta=-i\epsilon ,\]
 define a susy. transformation in superspace, where $\epsilon$ is
a complex Grasmannian parameter. The differential representation for the 
supercharge operators is given below: 

\begin{equation}  \label{cargadiferencial}
Q_{\bar{\theta }}=\partial _{\theta }+i\bar{\theta }\partial _{t},\qquad
\qquad Q_{\theta }=\partial _{\bar{\theta }}+i\theta \partial _{t},
\end{equation}
and the covariant derivatives are

\begin{equation}  \label{derivadacovariante}
D_{\bar{\theta }}=\partial _{\theta }-i\bar{\theta }\partial _{t},\qquad
\qquad D_{\theta }=\partial _{\bar{\theta }}-i\theta \partial _{t}.
\end{equation}
We have the following algebra,

\begin{equation}  \label{comutadordiferencial}
\left[ Q_{\bar{\theta }},Q_{\bar{\theta }}\right] _{+}=0,\quad \quad \left[
Q_{\theta },Q_{\theta }\right] _{+}=0,
\end{equation}

\[
\left[ Q_{\theta },Q_{\bar{\theta }}\right] _{+}=2i\frac{\partial }{\partial
t}=2H, 
\]
where $H$ will be identified with the Hamiltonian.

If we are able to define an invariant action, S, for a translation in
superspace $(t,\theta,\bar{\theta}) $ \cite{monopolospinsusyA},\cite
{capitulolivrosusy}, where the transformations parameter is a Grassmmanian
parameter, we say that this action has an N=2-supersymmetry. In this way,
we can define N=2-superfields \cite{susymecanicaquantica12} as we discuss in the sequel.

The complex chiral superfield ($\bar{D}\Phi_i=0$) admits the following $\theta$-expansion:

\begin{equation}
\Phi _{i}=x_{i}(t)+\theta \psi _{i}(t)+i\bar{\theta}\theta \dot{x}_{i}(t).
\label{campoquiral}
\end{equation}
$\psi _{i}$ is a Grasmannian variable and $x_{i}$ is a
commutative complex variable.

We have also a real superfield ($\Sigma=\Sigma^*$),

\begin{equation}
\Sigma =\xi(t)+i\theta \bar{\chi}(t)+i\bar{\theta}\chi(t) +\bar{\theta}\theta 
R(t) ,
\label{campoauxiliar}
\end{equation}
where $\xi $, $R$ are real component fields; $\chi $ and $\chi ^{\ast }$
are complex components fields; $\chi $ is a Grasmannian variable and $R$ is
a commutative variable. From the susy. transformations of these superfields, 
the component coordinates \cite{calculocarga} can be shown to transform as 
follows:

\begin{equation}  \label{transfromacoesdascomponentes}
\delta x_{i}=-i\epsilon \psi _{i}\quad \delta \psi _{i}=-2\epsilon ^{*}\dot{x%
}_{i}\quad \delta \xi =\epsilon ^{*}\chi +\epsilon \chi ^{*}
\end{equation}

\[
\delta \dot{x}_{i}=-i\epsilon \dot{\psi }_{i}\quad \delta \chi ^{*}=\epsilon
^{*}(R+i\dot{\xi })\quad \delta \chi =\epsilon (-R+i\dot{\chi }) 
\]

\[
\delta R=-i\epsilon ^{*}\dot{\chi }+i\epsilon \dot{\chi }^{*}.
\]
With these superfields, we define the N=2-action for a model similar to the previous
one by means of the expression,

\begin{equation}  \label{acaon2}
S=\int dtd\theta d\bar{\theta }{\cal {L}},
\end{equation}

\[
{\cal {L}}=\frac{i}{8}\frac{d\Phi _{i}}{dt}\Phi ^{*}_{i}+\frac{i}{8}\Phi _{i}%
\frac{d\Phi _{i}^{*}}{dt}+\frac{ie}{4}A_{i}(\Phi )\Phi _{i}^{*}-\frac{ie}{4}%
\Phi _{i}(A_{i}(\Phi ))^{*} 
\]

\[
-\frac{1}{4}(D_{\theta }\Sigma )(D_{\theta }\Sigma )^{*}+\frac{e\sqrt{2}}{4}%
\Sigma \Lambda (\Phi _{j})+\frac{e\sqrt{2}}{4}(\Lambda (\Phi ))^{*}\Sigma
^{*} 
\]

\begin{equation}  \label{lagrangeanan2}
+\frac{ie}{4}\epsilon _{ijk}\Lambda _{i}(\Phi )\Phi _{j}\Phi _{k}^{*}-\frac{%
ie}{4}\epsilon _{ijk}\Phi _{k}\Phi _{j}^{*}\Lambda _{i}(\Phi )^{*},
\end{equation}
where

\[
A_{i}(\Phi )=A_{i}(x)+\theta \psi _{j}\partial _{j}A_{i}(x)+i\theta
^{*}\theta \dot{x}_{j}\partial _{j}A_{i}(x),
\]

\[
\Lambda (\Phi )=\Lambda (x)+\theta \psi _{j}\partial _{j}\Lambda (x)+i\theta
^{*}\theta \dot{x}_{j}\partial _{j}\Lambda (x),
\]

\[
{\cal {B}}_{i}\equiv \Lambda _{i}\equiv \frac{\partial \Lambda (x)}{\partial
x_{i}}\qquad \Lambda _{ik}\equiv \frac{\partial ^{2}\Lambda (x)}{\partial
_{k}\partial x_{i}}\qquad A_{ji}\equiv \frac{\partial A_{j}(x)}{\partial
x_{i}}. 
\]

To work out the supersymmetric Lagrangian in terms of components, we start off from:

\[
{\cal {L}}=K+\theta V_{\bar{\theta }}+\bar{\theta }V_{\theta }+\bar{\theta }
\theta \widetilde{L}.
\]

Integrating eq. (\ref{acaon2}) on $\bar{\theta}$ and $\theta$, we get

\begin{equation}  \label{acaon2integrada}
S=\int dt {L},
\end{equation}
where we can write

\begin{equation}  \label{lagrangeanon2fisico}
\widetilde{L}=L+\frac{dY}{dt},
\end{equation}
the term $Y $ is a surface term, $\widetilde{L} $ is a total derivative of
surface term added to  $L $, and

\[
V_{\bar{\theta}}=-\frac{i}{4}\dot{x}_{i}\psi _{i}^{\ast }-\frac{i}{4}%
eA_{i}\psi _{i}^{\ast }+\frac{i}{4}ex_{i}(\partial _{j}A_{i})^{\ast }\psi
_{j}^{\ast }+\frac{i}{4}(R\chi +i\dot{\xi}\chi )+ 
\]

\[
\frac{i}{4}e\sqrt{2}\Lambda \chi -\frac{1}{4}e\sqrt{2}(\partial _{i}\Lambda
)^{\ast }\xi \psi _{i}^{\ast }+\frac{i}{4}e\sqrt{2}\Lambda ^{\ast }\chi -%
\frac{i}{4}e\epsilon _{ijk}\partial _{i}\Lambda x_{j}\psi _{k}^{\ast } 
\]

\begin{equation}
+\frac{i}{4}e\epsilon _{ijk}x_{k}x_{j}^{\ast }(\partial _{i}\partial
_{p}\Lambda )^{\ast }\psi _{p}^{\ast }+\frac{i}{4}e\epsilon
_{ijk}x_{k}(\partial _{i}\Lambda )^{\ast }\psi _{j}^{\ast }
\label{contribuiparacargan2}
\end{equation}

\[
V_{\theta }=\frac{i}{4}x_{i}^{\ast }\dot{\psi}_{i}-\frac{i}{4}eA_{i}^{\ast
}\psi _{i}+\frac{i}{4}ex_{i}^{\ast }\partial _{j}A_{i}\psi _{j}+\frac{i}{4}
(R\chi ^{\ast }-i\dot{\xi}\chi ^{\ast }) 
\]

\[
+\frac{i}{4}e\sqrt{2}\Lambda ^{\ast }\chi ^{\ast }+\frac{i}{4}e\sqrt{2}\Lambda \chi ^{\ast }-
\frac{i}{4}e\epsilon _{ijk}(\partial _{i}\Lambda )^{\ast }x_{j}^{\ast }\psi
_{k}+ 
\]

\[
\frac{1}{4}e\sqrt{2}\partial _{i}\Lambda \xi \psi _{i}
+\frac{i}{4}e\epsilon _{ijk}x_{k}^{\ast }x_{j}\partial _{i}\partial
_{p}\Lambda \psi _{p}+\frac{i}{4}e\epsilon _{ijk}x_{k}^{\ast }\partial
_{i}\Lambda \psi _{j}. 
\]

The Lagrangian in terms of the physical coordinates reads as below:

\[
L=\frac{1}{2}\dot{x}_{i}\dot{x}^{*}_{i}-\frac{i}{8}\dot{\psi }_{i}\psi
^{*}_{i}+\frac{i}{8}\psi _{i}\dot{\psi }^{*}_{i}+\frac{1}{2}e\, A_{i}\dot{x}%
_{i}^{*}-\frac{i}{4}e\, \partial _{j}A_{i}\psi _{j}\psi ^{*}_{i} 
\]

\begin{equation}  \label{lagrangeanon2fisicoaberto}
\frac{1}{2}e\, \dot{x}_{i}A^{*}_{i}+\frac{1}{4}ie\, (\partial
_{i}A_{j})^{*}\psi _{j}\psi ^{*}_{i}+\frac{1}{4}i\, \dot{\chi }\chi ^{*}-%
\frac{1}{4}i\, \chi \dot{\chi }^{*}
\end{equation}

\[
-\frac{1}{4}R^{2}-\frac{1}{4}\dot{\xi }^{2}+\frac{1}{4}e\, \sqrt{2}\left(
i\, \xi \dot{x}_{i}{\cal {B}}_{i}+i\, {\cal {B}}_{i}\psi _{i}\chi +R\Lambda
\right) 
\]

\[
+\frac{1}{4}e\, \sqrt{2}\left( \Lambda ^{*}R+i\, ({\cal {B}}_{i})^{*}\psi
^{*}_{i}\chi ^{*}-i\, ({\cal {B}}_{i})^{*}\dot{x}_{i}^{*}\xi \right) 
\]

\[
+\frac{1}{4}e\, \epsilon _{kji}\left( 2{\cal {B}}_{k}x_{j}\dot{x}%
_{i}^{*}-i\, {\cal {B}}_{k}\psi _{j}\psi ^{*}_{i}-i\, \partial _{r}{\cal {B}}%
_{k}x_{j}\psi _{r}\psi ^{*}_{i}\right) 
\]

\[
+\frac{1}{4}e\, \epsilon _{kji}\left( 2\dot{x}_{i}({\cal {B}}%
_{k})^{*}x^{*}_{j}+i\, ({\cal {B}}_{k})^{*}\psi _{i}\psi ^{*}_{j}+i\,
(\partial _{r}{\cal {B}}_{k})^{*}x^{*}_{j}\psi _{i}\psi ^{*}_{r}\right); 
\]
also, the surface terms collected in  $Y $ are as follows:

\begin{equation}  \label{termosdesuperficie}
Y=-\frac{1}{8}x_{i}\dot{x}^{*}_{i}-\frac{1}{8}\dot{x}_{i}x^{*}_{i}-\frac{e}{4%
}A_{i}x^{*}_{i}-\frac{e}{4}x_{i}A^{*}_{i}
\end{equation}

\[
-\frac{e}{4}\Lambda _{i}x_{j}x_{k}^{*}\epsilon _{ijk}-\frac{e}{4}%
x_{k}x_{j}^{*}\Lambda ^{*}\epsilon _{ijk} 
\]

\subsection{Supersymmetric charges from  the Lagrangian}

The Hamiltonian reads:

\[
H=\dot{x}_{i}\Pi _{x^{i}}+\Pi _{x^{i}}\dot{x}_{i}^{*}+\dot{\psi }_{i}\Pi
_{\psi _{i}}-\Pi _{\psi _{i}^{*}}\dot{\psi }^{*}_{i}+ 
\]

\[
+\dot{\chi }\Pi _{\chi }-\Pi _{\chi ^{*}}\dot{\chi }^{*}+\dot{\xi }\Pi _{\xi
}-L; 
\]
so,

\[
H=\left( \Pi _{x_{i}^{*}}-\frac{eA_{i}}{2}\right) \left( \Pi _{x_{i}}-\frac{%
eA^{*}_{i}}{2}\right) + 
\]

\begin{equation}  \label{hamiltonianan2}
+\frac{1}{8}ie\left( \partial _{j}A_{i}-(\partial _{i}A_{j})^{*}\right) \psi
_{j}\psi ^{*}_{i}+
\end{equation}

\[
\frac{i}{4}e\epsilon _{kji}{\cal {B}}_{k}\psi _{j}\psi _{i}^{*}+\frac{i}{4}%
e\epsilon _{kji}(\partial _{r}{\cal {B}}_{k})^{*}x_{j}^{*}\psi ^{*}_{r}\psi
_{i}+\frac{i}{4}e\epsilon _{kji}{\cal {B}}_{i}\chi \psi _{i} 
\]

\[
e\epsilon _{kji}\left( \Pi _{x_{i}^{*}}-\frac{eA_{i}}{2}\right) {\cal {B}}
^{*}_{k}x_{j}^{*}+\frac{e^{2}}{4}\epsilon _{kji}\epsilon _{npi}{\cal {B}}_{k}
{\cal {B}}^{*}_{i}x_{j}x_{p}^{*} 
\]

\[
\frac{i}{2}\sqrt{2}e\xi {\cal {B}}_{i}^{*}\left( \Pi _{x_{i}}-\frac{
eA^{*}_{i}}{2}\right) +\frac{e^{2}}{8}\xi ^{2}{\cal {B}}^{*}_{i}{\cal {B}}
_{i}-\frac{\Pi ^{2}_{\xi }}{2}-\frac{R^{2}}{8}+C.C., 
\]
where we take from the Lagrangian (\ref{lagrangeanon2fisicoaberto})  canonical
moments $\Pi _{j}$.

Using Noether's theorem for the supersymmetry transformations, we read the 
charges and we neglect the surface terms, which shall be considered in the 
next section. The idea is that here we are not interested in topological 
solutions.
\[
Q_{\bar{\theta }}=(\frac{5}{4}i\Pi _{x^{*}_{i}}-\frac{3}{8}ieA_{i}-\frac{5}
{16}e\sqrt{2}\xi \Lambda ^{*}_{i}+\frac{1}{8}iex_{j}A_{ij} 
\]

\begin{equation}  \label{cargan2semtermodesuperficie}
\frac{1}{8}ie\epsilon _{kij}x_{j}\Lambda ^{*}_{k}+\frac{1}{8}ie\epsilon
_{kri}x_{j}x_{r}\Lambda ^{*}_{kj}+\frac{1}{8}ieA^{*}_{ji}x_{j}
\end{equation}

\[
-\frac{1}{4}ie\epsilon _{jki}x_{k}\Lambda _{j}-\frac{1}{8}ie\epsilon
_{jrk}\Lambda ^{*}_{ji}x^{*}_{r}x_{k}+\frac{1}{4}ie\epsilon _{jrk}\Lambda
^{*}_{ij}x^{*}_{r}x_{k})\psi _{i}^{*} 
\]

\[
+(2\Pi _{\xi }+\frac{i}{2}e\sqrt{2}\Lambda ^{*}+\frac{i}{2}e\sqrt{2}\Lambda -
\frac{i}{8}e\sqrt{2}x_{i}\Lambda _{i})\chi 
\]

\[
Q_{\theta }=(-\frac{5}{4}i\Pi _{x_{i}}+\frac{3}{8}ieA_{i}^{*}-\frac{5}{16}e
\sqrt{2}\xi \Lambda _{i}-\frac{1}{8}iex^{*}_{j}A^{*}_{ij} 
\]

\[
-\frac{1}{8}ie\epsilon _{kij}x^{*}_{j}\Lambda _{k}-\frac{1}{8}ie\epsilon
_{kri}x^{*}_{j}x^{*}_{r}\Lambda _{kj}-\frac{1}{8}ieA_{ji}x^{*}_{j}+ 
\]

\[
\frac{1}{4}ie\epsilon _{jki}x_{k}\Lambda _{j}+\frac{1}{8}ie\epsilon
_{jrk}\Lambda _{ji}x_{r}x^{*}_{k}-\frac{1}{4}ie\epsilon _{jrk}\Lambda
_{ij}x_{r}x^{*}_{k})\psi _{i} 
\]

\[
+(2\Pi _{\xi }-\frac{i}{2}e\sqrt{2}\Lambda -\frac{i}{2}e\sqrt{2}\Lambda ^{*}+
\frac{i}{8}e\sqrt{2}x^{*}_{i}\Lambda ^{*}_{i})\chi ^{*}. 
\]

We check that

\begin{equation}  \label{algebran2cargassemtermosdesuperficie}
\left[ Q_{\bar{\theta }},Q_{\theta }\right]_{+} =2H_{1}
\end{equation}

\[
\left[ Q_{\theta },Q_{\theta }\right]_{+} =\frac{5}{32}%
iex_{k}^{*}A_{k[j,i]}-\left( \frac{5}{8}\right) ^{2}e\sqrt{2}\xi \Lambda
_{[i,j]} 
\]

\[
-\frac{5}{16}ie\Lambda _{[i,k]}x^{*}_{n}\epsilon _{kjn}-\frac{5}{16}%
ie\Lambda _{[k,j]}x^{*}_{n}\epsilon _{kin} 
\]

\[
-\frac{5}{32}iex^{*}_{k}x_{r}\epsilon _{nrk}(\Lambda _{nji}-2\Lambda
_{jni}-\Lambda _{nij}+2\Lambda _{inj})\psi _{i}\psi _{j} 
\]

\[
\left[ Q_{\bar{\theta }},Q_{\bar{\theta }}\right]_{+} =(-\frac{5}{32}%
iex_{k}A^{*}_{k[i,j]}-(\frac{5}{8})^{2}e\sqrt{2}\xi \Lambda
^{*}_{[j,i]} 
\]

\[
+\frac{5}{16}ie\Lambda ^{*}_{[j,k]}x_{n}\epsilon _{kin}+\frac{5}{16}%
ie\Lambda ^{*}_{[k,i]}x_{n}\epsilon _{kjn}+ 
\]

\[
\frac{5}{32}iex_{k}x^{*}_{r}\epsilon _{nrk}(\Lambda ^{*}_{nij}-2\Lambda
^{*}_{inj}-\Lambda ^{*}_{nji}+2\Lambda ^{*}_{jni})\psi ^{*}_{j}\psi ^{*}_{i}, 
\]
where the subscript $[i,j]$ stands for anti-symmetrisation over the indices i,j.
If we impose that the algebra of charges are the ones in (\ref
{comutadordiferencial}), we need that

\begin{equation}  \label{condicaoparaalgebran2semcargasuperficie}
A_{k[j,i]}\equiv 0\qquad \Lambda _{[i,j]}\equiv 0\qquad \Lambda
_{r[j,i]}\equiv 0
\end{equation}

\[
A^{*}_{k[j,i]}\equiv 0\qquad \Lambda ^{*}_{[i,j]}\equiv 0\qquad
\Lambda ^{*}_{r[j,i]}\equiv 0 ,
\]
which means that the funtions have continuous second derivatives. We then 
see that central charges in the N=2 - su.sy. algebra might appear only if we 
have potential configurations with non-continuous second derivatives, like
what happens in the presence of Dirac-string monopoles.

\subsection{Supersymmetric charges from the Lagrangian with the surface terms}

If we now consider the surface terms coming from $Y $, We get to the
Lagrangian

\[
\widetilde{L}=L+\frac{dY}{dt}. 
\]
 
We now consider the contributions arising from the surface terms, since 
they might lead to topological configurations that could induce the 
appearance of central charges in the N=2-susy. algebra.

\[
Q_{\bar{\theta }}=(\frac{4}{3}i\Pi _{x^{*}_{i}}-\frac{1}{3}ieA_{i}-\frac{1}{3%
}e\sqrt{2}\xi \Lambda ^{*}_{i}+\frac{1}{3}iex_{j}A_{ji} 
\]

\[
\frac{1}{3}ie\epsilon _{jik}x_{k}\Lambda ^{*}_{j}+\frac{1}{3}ie\epsilon
_{rjk}x_{k}x_{j}^{*}\Lambda ^{*}_{ri}-\frac{1}{3}ie\epsilon
_{jki}x_{k}\Lambda _{j})\psi _{i}^{*} 
\]

\begin{equation}  \label{caargan2comtermodesuperficie}
+(2\Pi _{\xi }+\frac{i}{2}e\sqrt{2}\Lambda ^{*}+\frac{i}{2}e\sqrt{2}\Lambda
)\chi
\end{equation}

\[
Q_{\theta }=(-\frac{4}{3}i\Pi _{x_{i}}+\frac{1}{3}ieA^{*}_{i}-\frac{1}{3}e%
\sqrt{2}\xi \Lambda _{i}-\frac{1}{3}iex^{*}_{j}A^{*}_{ji} 
\]

\[
-\frac{1}{3}ie\epsilon _{jik}x^{*}_{k}\Lambda _{j}-\frac{1}{3}ie\epsilon
_{rjk}x^{*}_{k}x_{j}\Lambda _{ri}+\frac{1}{3}ie\epsilon
_{jki}x^{*}_{k}\Lambda ^{*}_{j})\psi _{i} 
\]

\[
(2\Pi _{\xi }-\frac{i}{2}e\sqrt{2}\Lambda -\frac{i}{2}e\sqrt{2}\Lambda
^{*})\chi ^{*} 
\]

following that the charges algebra are,

\[
\left[ Q_{\bar{\theta }},Q_{\theta }\right]_{+} =2\widetilde{H} 
\]

\[
\left[ Q_{\theta },Q_{\theta }\right]_{+} =(\frac{4}{9}eix^{*}_{k}A_{k[j,i]}+%
\frac{4}{9}e\sqrt{2}\xi \Lambda _{[i,j]} 
\]

\begin{equation}  \label{comutadorcargan2comtermodesuperficie}
+\frac{4}{9}eix_{n}x_{k}^{*}\epsilon _{rnk}\Lambda _{r[j,i]})\psi
_{i}\psi _{j}
\end{equation}

\[
\left[ Q_{\bar{\theta }},Q_{\bar{\theta }}\right]_{+} =(-\frac{4}{9}%
eix_{k}A^{*}_{k[i,j]}+\frac{4}{9}e\sqrt{2}\xi \Lambda ^{*}_{[j,i]} 
\]

\[
-\frac{4}{9}eix_{n}^{\ast }x_{k}\epsilon _{rnk}\Lambda _{r[i,j]}^{\ast
})\psi _{j}^{\ast }\psi _{i}^{\ast }, 
\]
where the Hamiltonian $\widetilde{H}$ is the same obtained from lagrangian $
\widetilde{L}$ by Legendre transformation.

If we impose that the algebra of charges is the same as the one in (\ref
{comutadordiferencial}), we need that

\begin{equation}  \label{condicaoparaalgebran2comtermosuperficie}
A_{k[j,i]}\equiv 0\qquad \Lambda _{[i,j]}\equiv 0\qquad \Lambda
_{r[j,i]}\equiv 0
\end{equation}

\[
A^{*}_{k[j,i]}\equiv 0\qquad \Lambda ^{*}_{[i,j]}\equiv 0\qquad
\Lambda ^{*}_{r[j,i]}\equiv 0; 
\]
the algebra of commutation relations for  the component coordinates are 
given as:

\[
\left[ x_{i},\Pi _{x_{k}}\right] _{-}=i\delta _{ik}\quad \left[
x^{*}_{i},\Pi _{x_{k}^{*}}\right] _{-}=i\delta _{ik}\quad \left[ \xi ,\Pi
_{\xi }\right] _{-}=i 
\]

\begin{equation}  \label{comutadorescomponentesn2comtermosuperficie}
\left[ \psi _{i},\psi _{k}^{*}\right] _{+}=8\delta _{ik}\qquad \left[ \chi
,\chi ^{*}\right] _{+}=-4,
\end{equation}

\[
\left[ \psi _{i},\Pi _{\psi _{k}}\right] _{+}=-i\quad \left[ \psi
^{*}_{i},\Pi _{\psi ^{*}_{k}}\right] _{+}=-i\quad \left[ \chi ,\Pi _{\chi
}\right] _{+}=-i 
\]

\[
\left[ \chi ^{\ast },\Pi _{\chi ^{\ast }}\right] _{+}=-i ;
\]
all other relations vanish. We can obtain the same algebra for the
components of superfields, proceeding to the canonical quantization \cite
{quantizacaosusyvinculo}, where the momenta $\widetilde{\Pi }_{j}$, follow 
now from the Lagrangian $\widetilde{L}$.

\section{Conclusions}

We propose a supersymmetric quantum-mechanical system given by a 
non-relativistic particle in an external field. We formulate the problem in 5D, 
where the
particle presents 4 fermionics degrees of freedom. and 4 bosonics 
degrees of freedom (N=1 - D=4 $=>$ N=2 - D=4). In dimensions smaller than 3+1, 
we do not have suficient degrees of freedom in the fermionic sector. 
Then, we cannot write the model in terms
of superfields.

Starting from 4+1, the consequence is that we drag another object from there,
${\cal{B}}_{i}$. The interpretation of this object is not clear, but it
follows from the extended Electromagnetism we have discussed. We obtain a
simple and also extended (N=2) su.sy..

We observe that, although the charge in N=2-su.sy. changes with the surface
terms in the Lagrangian, we do not have change the charges algebra of
supersymmetry. In other words, surface terms do not give rise to a
central charge in this model, except in the case the derivatives do not
commute when applied on $A^{\mu}$ and $\Lambda$, or, in other words, when
Dirac-like strings are present.

Here, we have set an extended eletromagnetic model to see how central charges may
show up. Next, we wish to reset the turbulent system, decribed by a 
BRS-symmetry, as studied in \cite{aplicacaosusyg} in terms of an 
N=2-D=2-supersymmetry. The instanton solution present in the system may be the
source of a central charge in the N=2-algebra. Also, the realisation of an 
N=2-su.sy. for the turbulent system may ensure the topological stability of
dual solutions that saturate the Bogomol'nyi bound 
\cite{susymecanicaquantica12},\cite{shapeinvariance}. 

\section*{Appendix}
We quote below the  $\Gamma$-matrices representing the Clifford algebra of 
(1+4)D:
\[\Gamma ^{5}=\left( 
\begin{array}{cc}
0 & 1_{2} \\ 
1_{2} & 0
\end{array}
\right);\quad \Gamma ^{i}=\left( 
\begin{array}{cc}
0 & \sigma ^{i} \\ 
-\sigma ^{i} & 0
\end{array}
\right);\quad 
\Gamma ^{0}=\left( 
\begin{array}{cc}
1_{2} & 0 \\ 
0 & -1_{2}
\end{array}
\right);\quad \Gamma ^{4}=-i\Gamma ^{5}=\left( 
\begin{array}{cc}
0 & i_{2} \\ 
i_{2} & 0
\end{array}
\right). \]

\section*{ACKNOWLEDGEMENTS}

The authors are grateful to A. L. M. A. Nogueira and 
W.A. Moura-Melo for pleasant and clarifying discussion. M. Plyushchay is also 
acknowledged for the interest in pointing out a number of relevant references. 
The authors express their gratitude to CNPq for the financial support

\end{document}